\documentclass[preprint,12pt,numbers]{elsarticle}

\usepackage{amssymb}
\usepackage{amsmath}
\usepackage{graphicx} 
\usepackage{booktabs}
\usepackage{float}
\journal{Superconductivity}

\begin{document}

\begin{frontmatter}



\title{Detector Development for HUBS I: Initial Testing of Small-Area TES Microcalorimeters}


\author{Naihui Chen}
\author{Jian Ma}
\author{Sifan Wang\corref{*}}
\author{Qian Wang}
\author{Guanhua Gao}
\author{Qing Yu}
\author{Yajie Liang}
\author{Yaowu Song}
\author{Jingyi Zhang}
\author{Rui Huang}
\author{Wei Cui\corref{*}} %% Author name

\cortext[*]{To whom correspondence should be addressed. Email: sifanwang@tsinghua.edu.cn, cui@tsinghua.edu.cn}

\affiliation{organization={Low Temperature Detector Lab, Department of Astronomy, Tsinghua University},%Department and Organization
            city={Beijing},
            postcode={100084}, 
            country={China}}

%% Abstract
\begin{abstract}
We report progress on the ongoing development of microcalorimeter detector technology for the Hot Universe Baryon Surveyor (HUBS) mission. We show the results from testing and characterizing selected pixels in a 10$\times$10 microcalorimeter array. The microcalorimeter is based on a Mo/Cu transition-edge sensor (TES) coupled to an Au absorber. To better understand the properties of the devices, we have first measured the energy resolution of a selected pixel in a TES array of the same design with a pulsed laser system that produces 3 eV photons, and found that individual photon peaks are easily resolved with the TES, indicating good performance.
We have then exposed the microcalorimeter array to radiation from a $^{55}$Fe source, and found that the pixels tested show energy resolutions as good as $3.7\pm0.1~\mathrm{eV}$ at $5.9~\mathrm{keV}$. The energy resolution is found to vary monotonically with the bias point for all the devices, showing little evidence for the presence of the so-called excess noise. This is consistent with the results from modeling the measured noise spectrum. The effects of thermal crosstalk are evident, leading to the degradation of energy resolution.

\end{abstract}



%% Keywords

\end{frontmatter}



\section{Introduction}

The Standard Cosmological Model (also known as the $\Lambda$CDM model) attributes about 5\% of the energy density of the present-day universe to normal matter (also known as baryonic matter), which is the observable, and the rest to new physics (specifically dark matter and dark energy).
The results from optical surveys indicate that, as the universe evolves, more and more cosmic baryons seem to go ``missing'', with only about half are detected in the present-day universe, leading to the long-standing ``missing baryon problem'' \cite{fukugita1998cosmic}. On the theoretical front, cosmological simulations suggest that, during the formation of structures, some of the baryons may be heated to about $10^6~\mathrm{K}$ and be present in a gaseous phase of very low density. Such hot baryons would radiate very weakly in the soft X-ray band, not detectable with current observing facilities and thus ``missing''.  

A non-dispersive, high-resolution X-ray spectrometer would make it possible to detect the ``missing'' baryons \cite{cui2020hubs}, through narrow-band imaging and X-ray spectroscopy, because the spectrum of their radiation is expected to be dominated by emission lines. To this end, the HUBS project has been proposed \cite{cui2020hubs2}. HUBS will be equipped with an array of microcalorimeters.
A microcalorimeter consists mainly of three components \cite{mccammon1984experimental}: an effective X-ray absorber, a sensitive temperature transducer, and a weak thermal link. The energy of an incident X-ray photon is deposited in the absorber and is thermalized, causing an increase in the temperature of the device. The temperature transducer senses the temperature change and converts the thermal signal to an electrical signal that is read out. After the thermal energy is transferred to the heat sink through the thermal link, the temperature of the device recovers, and it is ready for the arrival of the next photon. 

Therefore, every X-ray photon produces a pulse signal, with its amplitude proportional to the energy of the incident photon. The accuracy in measuring the pulse amplitude, in the presence of noise, determines the energy resolution of the microcalorimeter.
Often, a thermistor is chosen as the temperature transducer. In this case, the theoretical limit of energy resolution is expressed by \cite{mccammon2005thermal} 
\begin{equation}
\Delta E = \xi \sqrt{\frac{k_{\rm B} T_0^2 C}{\alpha}}, \label{Eresolution}
\end{equation}
where $k_{\rm B}$ is the Boltzmann constant, $\xi$ a dimensionless factor, $T_0$ the microcalorimeter temperature, $C$ the heat capacity, and $\alpha = \partial\log R / \partial \log T$ the temperature sensitivity of the thermistor. Not as explicit is the current sensitivity of the thermistor, $\beta = \partial \log R / \partial \log I$, which also affects the energy resolution.

The challenge for developing HUBS detector lies in the difficulty of balancing the need for both large-area pixels and high energy resolution \cite{wang2022development}. In this work, however, we present results from testing initial batches of $10\times 10$ microcalorimeter arrays of smaller-area pixels. The temperature transducer is based on the superconducting transition-edge sensor (TES) technology \cite{irwin2005transition}. The TES is made of a Mo/Cu bilayer film, and the absorber is made of pure Au and of area $240\times 240$ $\mu \rm m^2$. 

\section{Device Fabrication}
\begin{figure}[H]
\centering
\includegraphics[height = 5 cm]{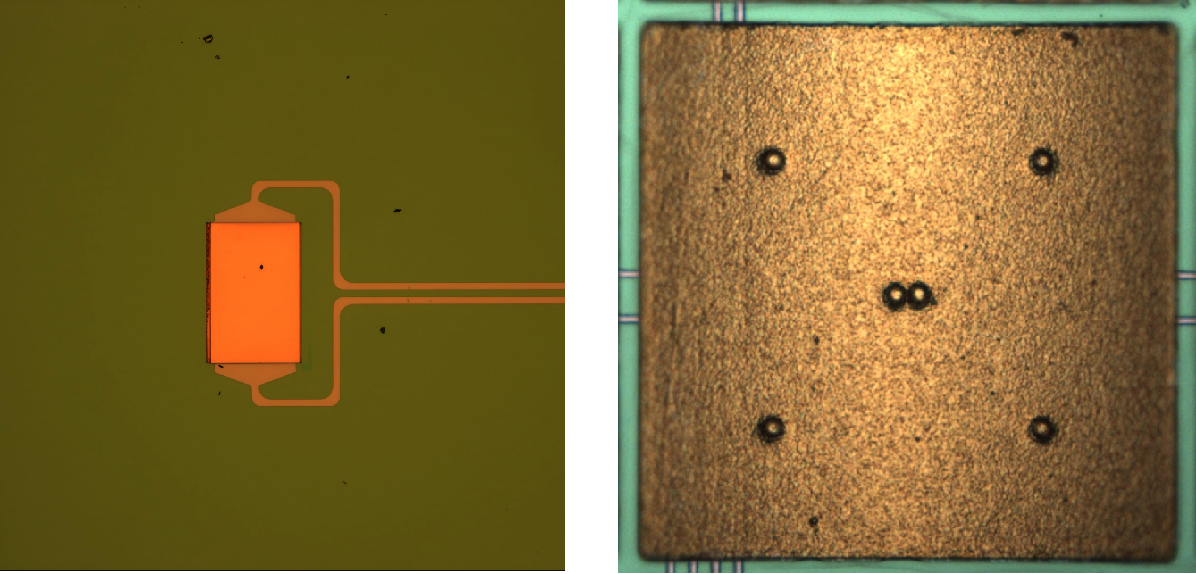}
\caption{\label{fig:sample} Left: Micrograph of the Mo/Cu TES with Mo leads. The size of the TES is $120~\mu\text{m}\times80~\mu\text{m}$. Right: Micrograph of the microcalorimeter with an Au absorber, which is connected to the TES via two central stems. The size of the TES in this microcalorimeter is $60~\mu\text{m}\times20~\mu\text{m}$, and the leads are made of Nb. The size of the absorber is $240~\mu\text{m}\times240~\mu\text{m}$.}
\end{figure}

The fabrication of the detector array basically follows the process described in \cite{wang2022development}. 
The base detector material is a Mo/Cu bilayer film \cite{wang2022-sust}. The bilayer film is deposited, via DC magnetron sputtering, on a Si wafer with both sides covered with low-stress ${\rm SiO}$ and ${\rm SiN}$ thin films. The thickness of the Mo or Cu layer is fine-tuned so that the superconducting transition temperature ($T_{\rm c}$) of the bilayer is around $80~\mathrm{mK}$. The bilayer film is patterned and processed through a number of lithography and etching steps, to produce Mo/Cu TES arrays of varying designs. The edge of a TES is specially treated to significantly enhance the steepness of the superconducting transition \cite{wang2022-sust}. 
The Au absorbers are then added to the TES array via electroplating. In this work, the size of the absorber is $240~\mu\text{m}\times240~\mu\text{m}$, and the thickness is around $2~\mu\text{m}$. The absorber is supported by six stems, with four corner ones on ${\rm SiN}$ and two central ones on TES (see Fig. \ref{fig:sample}). We note that the devices tested here have not undergone Si back-side etching. 

\section{Experiment Setup}
%\subsection{Measurement System}
The measurements are carried out in a dilution refrigerator, with a setup that is similar to the one described in \cite{wang2022development}. 
% As shown in Fig. \ref{fig:setup} t
The sample box is mounted on the mixing‑chamber stage and cooled below 100 mK. The device under test is read out with an input SQUID and a SQUID array, which are mounted on the cold stage, followed by a room‑temperature flux‑locked loop \cite{clarke2006squid} for linearized signal conditioning. A voltage bias is applied through a small shunt resistor to operate the TES in a negative electrothermal-feedback mode.

% \section{Device Characterization}
\section{TES Characterization}
\subsection{$I$‑$V$ Measurements}
\label{sec:tesIV}
\begin{figure}[H]
    \begin{center}
    \includegraphics[width=1.0\linewidth]{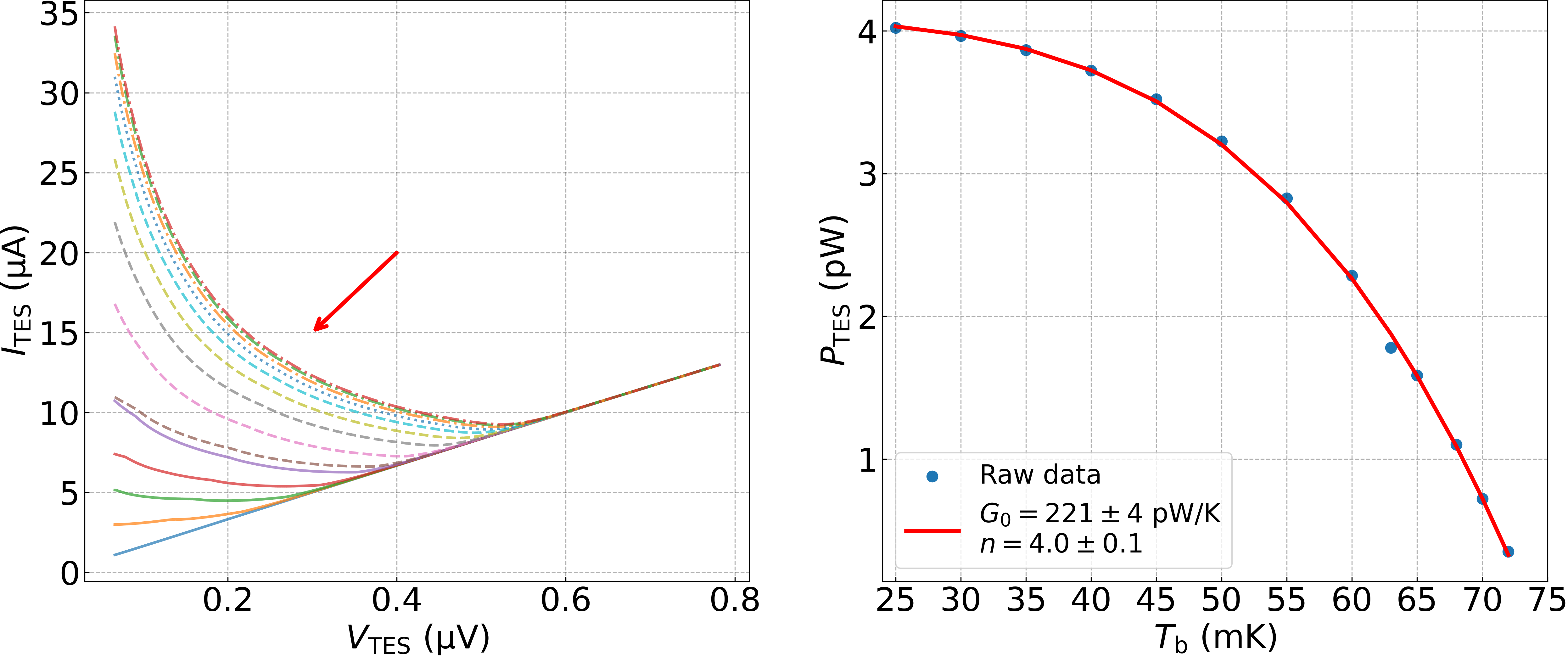}% Here is how to import EPS art
    \caption{\label{fig:IV} Electrical and thermal properties of the tested TES. Left: Measured $I$–$V$ curves. Along the direction of the (red) arrow, $T_\mathrm{b}$ increases from $25~\mathrm{mK}$ to $75~\mathrm{mK}$. Right: Measured $P$–$T$ curve. It is derived from the $I$–$V$ curves, with the best-fit model (see the main text) shown in solid (red) line.
}
    \end{center}
\end{figure}

We have obtained the $R$-$T$ curve of a TES pixel that is selected from a 10$\times$10 array, and found that
%using a four‑probe lock-in measurement. 
its normal-state resistance ($R_{\rm n}$) is 60 m$\Omega$. 
By sweeping the bias voltage, we have measured its $I$‑$V$ characteristics, from the normal state to the superconducting state, and the results are shown in Fig. \ref{fig:IV}, for different base temperatures of the heat sink, from which a relationship between power dissipation and base temperature ($P$-$T$ curve) is derived. Assuming the thermal conductance of the weak link is a power-law function of temperature,
\begin{equation}
G = G_0 \left( \frac{T}{T_{\rm c}} \right)^{n-1},
\end{equation}
we have
\begin{equation}
P = \int_{T_{\rm b}}^{T_{\rm c}} G{\rm d}T=\frac{G_0}{nT_{\rm c}^{n-1}} (T_{\rm c}^n - T_{\rm b}^n), \label{GTfit}
\end{equation}
where $T_{\rm b}$ is the base temperature, and the TES sits at $T_{\rm c}$. Applying it to the measured $P$-$T$ curve of the device, we show the best fit in Fig. \ref{fig:IV} (right panel), with $G_0 = 221~\mathrm{pW/K}$, $T_{\rm c} = 73.5~\mathrm{mK}$ and $n = 4.0$ being the corresponding values.

\subsection{Optical Illumination Measurements} \label{sec:optical}
Because heat capacity is kept small for TES devices, they are easily saturated when illuminated by an X-ray source. To characterize the energy resolution, we used a pulsed laser system, following \cite{jaeckel2019energy}. In this setup, each pulse (containing $N$ 3 eV photons) is used to simulate an X-ray photon (of energy 3$N$ eV), when the duration of the pulse is much shorter than the response time of the devices. A monochromatic light source of wavelength $405~\mathrm{nm}$ is located outside the refrigerator, and is driven by a waveform generator, allowing the pulse duration and optical power to be adjustable.
The light enters the refrigerator through an optical fiber, which terminates just in front of the device under test.
To suppress thermal crosstalk caused by photons hitting the ${\rm SiN}$ outside the TES, a $100~\mu$m-thick mask with $80~\mu$m-diameter apertures was positioned $0.3~\mathrm{mm}$ above the device.

\begin{figure}[H]
\begin{center}
\includegraphics[width=1\linewidth]{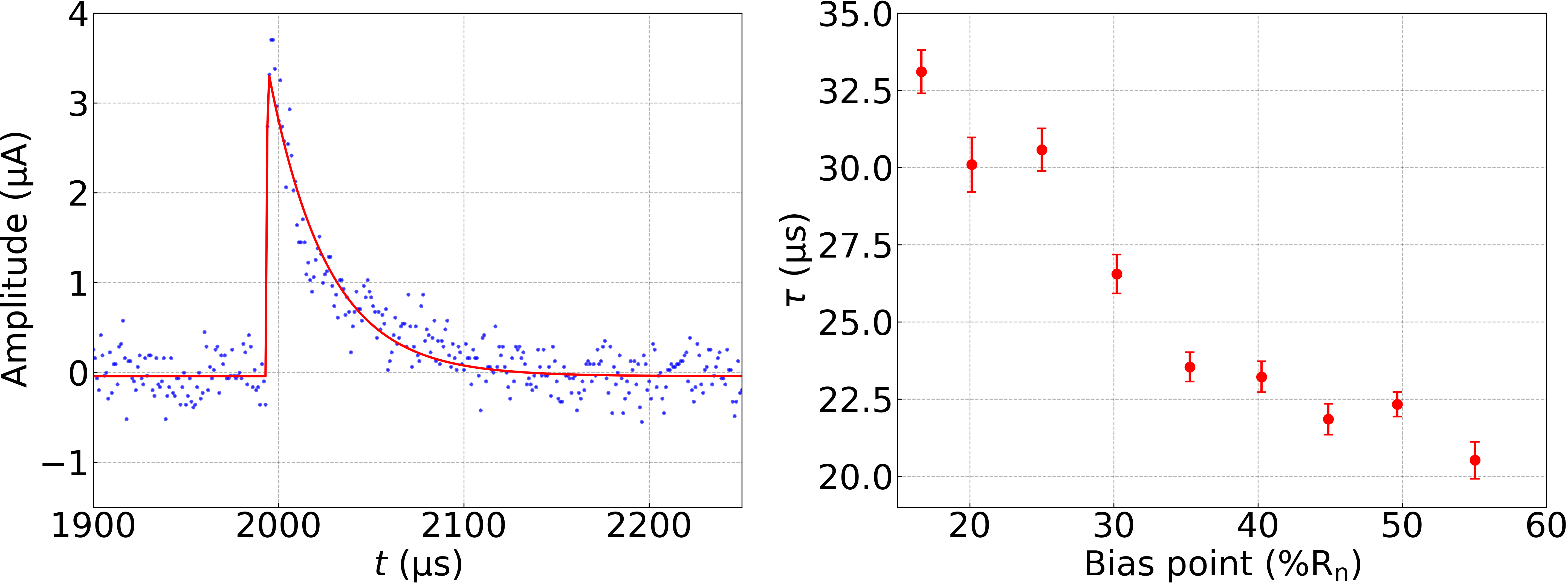}
\caption{\label{fig:tau} TES response to illuminating laser pulse. Left: Average pulse profile. The best-fit double-exponential function is shown in solid (red) line.
%Dots represent the measured data, and the curve shows the fitting result.
Right: Pulse decaying time constant as a function of the TES bias point.
}
\end{center}
\end{figure}

For a given bias point, the average profile of TES response to laser pulses is shown in Fig. \ref{fig:tau} (left panel). We use a double exponential function with rising and decaying time constants to fit the profile, and the best-fit model is also shown in the figure. 
The decaying time constant decreases with the bias point of the device, ranging roughly from $20$ to $30~\mu\text{s}$, as shown in the right panel of the figure.

To determine the intrinsic energy resolution of the TES, we have lowered the intensity of the input laser light to enter single-photon mode. 
The pulse heights are estimated using an optimal filter (OF) algorithm \cite{szymkowiak1993signal}.
Since the pulse arrival time is random with respect to the sampling clock, we extend the standard
frequency-domain OF with a phase-correction term to compensate for the resulting sub-sample time
offset $\tau$:

\begin{equation}
\hat{E} = \frac{\displaystyle
\sum_k \frac{1}{J(f_k)}\;
\operatorname{Re}\!\bigl[ D(f_k)\,S^{\ast}(f_k)\,e^{+j\omega_k\hat{\tau}} \bigr]\,\Delta f}
{\displaystyle
\sum_k \frac{1}{J(f_k)}\;|S(f_k)|^2\,\Delta f},
\qquad \omega_k \equiv 2\pi f_k.
\end{equation}
The time offset $\hat{\tau}$ is obtained by convolving the recorded pulse with the optimal filter
kernel $\mathrm{OF}(t) \equiv \mathcal{F}^{-1}[S^{\ast}(f)/J(f)]$ and locating the peak of the
convolution curve \cite{szymkowiak1993signal}. A detailed derivation is given in \ref{app1}. 

\begin{figure}[H]
    \begin{center}
    \includegraphics[width=1\linewidth]{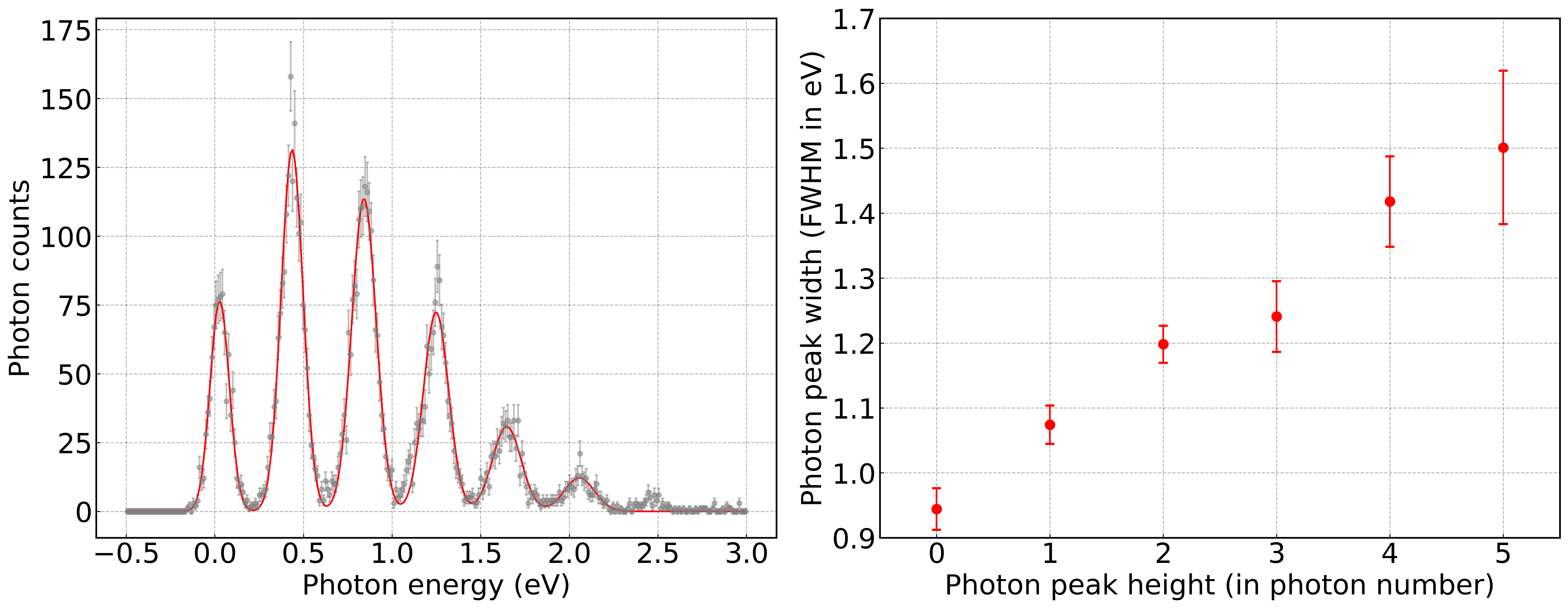}
    \caption{\label{fig:Eresolution} TES response to monochromatic light pulses, when biased at 25\% $R_{\rm n}$. Left: Pulse height distribution. The data are acquired at a constant laser power, and the mean photon number per pulse is determined to be around 2. The pulse heights are calibrated based on the separation of adjacent photon peaks (3 eV), and converted to photon energies. The best-fit model is shown in solid (red) line.
    %Dots represent the raw data and the curve shows the fitting result. 
    Right: Energy resolution (FWHM). The results are shown for different photon peaks. }
    \end{center}
 
\end{figure}

Individual photon peaks are clearly identified in the pulse-height distribution, as shown in Fig. \ref{fig:Eresolution} (left panel). Here we note that these data were taken at a fixed laser power, and the mean photon number per pulse at this laser power was around $2$. Fitting the distribution with a series of Gaussian functions (with the centroids fixed at the respective photon energies) yields the resolution of the TES at different photon peaks. From the zero‑photon peak we extract %intrinsic single-photon energy resolution of the device 
the corresponding energy resolution as
$\Delta E= 0.94\, \pm \, 0.04 $  eV (FWHM), which includes contributions from readout noise (and environmental noises), as well as from ``dark'' detector noise.
Fig. \ref{fig:Eresolution} also shows the measured resolution for the first five photon peaks (in the right panel). The degradation in the resolution is apparent towards higher photon numbers.

\section{Microcalorimeter Characterization}
We have selected two $10\times10$ microcalorimeter arrays for characterization, which adopt the same TES design as the one tested (see previous section), but are fabricated on a different batch of substrate wafers (see discussion below). The absorbers are made of pure gold and have an area of $240~\mu\text{m} \times 240~\mu\text{m}$.

\subsection{Sample Devices}
We have picked five devices for measurements, Devices S3-4 and S4-4 from one array (chip D3) and the remaining three devices from the other (chip D2). Four of the devices are tested with a $^{55}$Fe source, as indicated in Tab. \ref{tab:ucal}, and two are characterized electrically when shielded from the source (``Dark Test''). The measurements are affected by different SQUIDs used in the readout electronics.  
For SQUIDs with a coupling ratio ($\frac{1/M_{\rm fb}}{1/M_{\rm in}}$, where $M_{\rm fb}$ and $M_{\rm in}$ are mutual inductances with feedback coil and input coil, respectively) close to 12, we are not able to maintain stable operation at low bias points (below 20\% $R_{\rm n}$), so the corresponding results are missing.

\begin{table}[htbp]
    \centering
    \caption{Information of measured microcalorimeters and corresponding SQUID coupling coefficients.}
    \label{tab:ucal}
    \begin{tabular}{@{}c c c c c c@{}}  % @{} 消除两侧多余空白
        \toprule
        Chip & $\mu$Cal pixel & TES size ($\mu\text{m}^2$) & $^{55}$Fe test & Dark test & $\frac{1/M_{\rm fb}}{1/M_{\rm in}}$ \\
        \midrule
        D3 & S3-4 & $20 \times 60$ & No & Yes & 2.85 \\
        D3 & S4-4 & $20 \times 60$ & Yes & Yes & 2.85 \\
        D2 & S2-5 & $20 \times 50$ & Yes & No & 2.86 \\
        D2 & S3-5 & $20 \times 50$ & Yes & No & 12.77 \\
        D2 & S4-5 & $20 \times 50$ & Yes & No & 12.68 \\
        \bottomrule
    \end{tabular}
\end{table}

\subsection{$I$-$V$ Measurements}
$I$-$V$ characterization is carried out on each of the selected microcalorimeters for different base temperatures, and thermal conductance is then characterized through $P$-$T$ curves.
Tab. \ref{tab:iv_params} summarizes the results.

\begin{table}[htbp]
    \centering
    \caption{Summary of $I$-$V$ measurements of microcalorimeters.}
    \label{tab:iv_params}
    \begin{tabular}{@{}c c c c c c c@{}}  % @{} 消除两侧多余空白
        \toprule
        Chip & $\mu$Cal pixel & $T_{\rm c}$ (mK) & $R_{\rm n}~\rm (m\Omega)$ & $G_0$ (pW/K) & $n$ & $K_{\rm Kap}$ $\rm(W/K^4/m^2)$ \\
        \midrule
        D3 & S3-4 & 87 & 103 & 84 & 4.6 & 26.6\\
        D3 & S4-4 & 87 & 104 & 92 & 4.7 & 29.1\\
        D2 & S2-5 & 83 & 82 & 58 & 4.6 & 25.4\\
        D2 & S3-5 & 84 & 91 & 52 & 4.4 & 21.9\\
        D2 & S4-5 & 83 & 88 & 49 & 4.4 & 21.4\\
        
        \bottomrule
    \end{tabular}
\end{table}

The measured thermal conductance $G_0$ of the absorber-coupled TES devices is significantly lower than that of the bare TES device (see Sec. \ref{sec:tesIV}). Given $n \sim 4$ in all cases, Kapitza boundary resistance seems to be dominating \cite{swartz1989thermal}. The Kapitza constant $K_{\rm Kap}$ ($= G_0 / (4 A T_{\rm c}^{3})$, where $A$ is the area of the Kapitza interface, i.e., the TES area in this case) is 14.5 $\rm W/K^4/m^2$ for the bare TES device, which is much smaller than that for the absorber-coupled TES devices (24.9 $\rm W/K^4/m^2$ on average). The difference in $K_{\rm Kap}$ is also apparent between the two arrays of microcalorimeters (Chip D3 and D2). These discrepancies might be related to non-uniformity between different fabrication batches and within the same batch, including SiN/SiO and TES bilayer films. We note that $K_{\rm Kap}$ in this work is significantly lower than other reported values, e.g., 125 $\rm W/K^4/m^2$ in \cite{HOEVERS_2006}.

\subsection{X-ray Illumination Measurements}

The $^{55}$Fe source is placed about
%in the holder of the lid directly above the microcalorimeter, at a distance of 
9 $\mathrm{mm}$ from a microcalorimeter array under test. 
No pinhole was used, so all pixels in the array are irradiated at the same time. For absorbers of area $240~\mu\text{m}\times240~\mu\text{m}$, the expected event rate is approximately 1.2 counts/s/pixel.

\begin{figure}[ht]
\centering
\includegraphics[width=\linewidth]{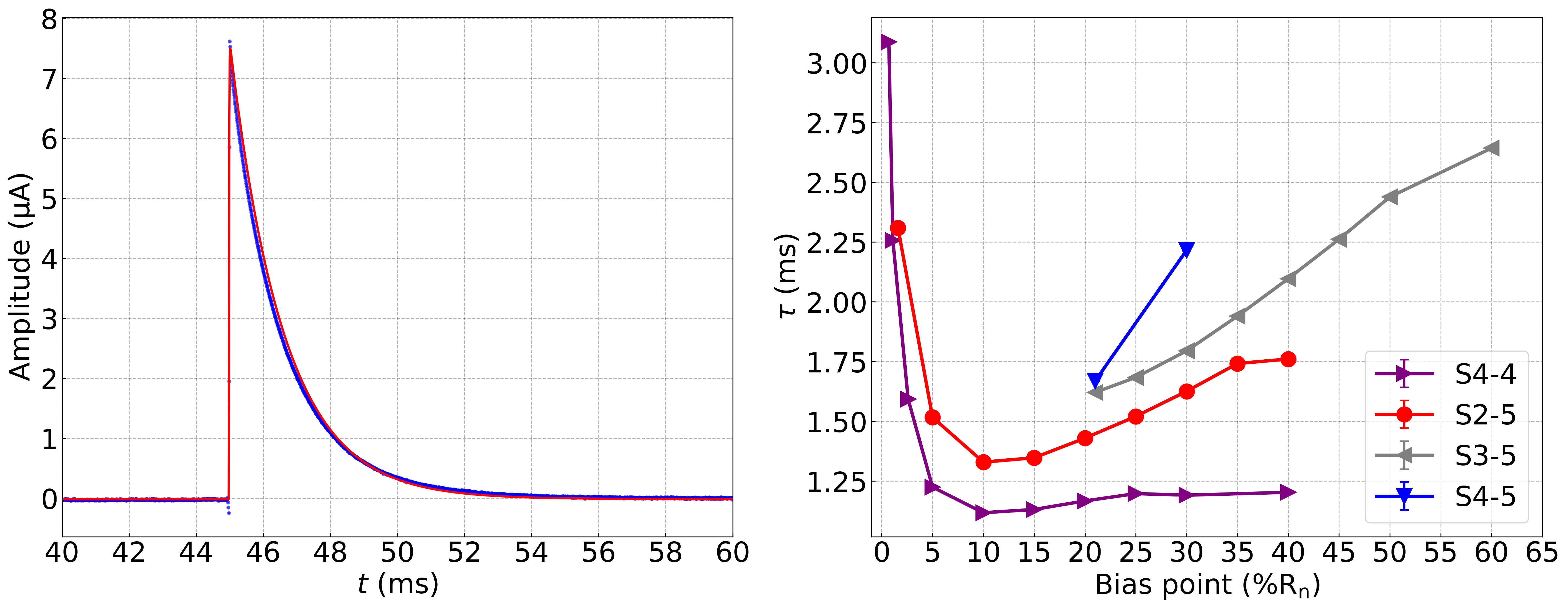}% Here is how to import EPS art
\caption{\label{fig:tau_Tb} Results from $^{55}$Fe illumination. Left: Average X-ray pulse profile. The best-fit double-exponential function is shown in solid (red) line.  
Right: Pulse decaying time as a function of bias. The results for all devices are shown. Note that no results at low bias points are shown for two of the devices due to issues with the SQUIDs used in the readout electronics (see the main text).
}
\end{figure}

Fig. \ref{fig:tau_Tb} shows the average profile of X-ray pulses (in the left panel). We fit the profile with a double-exponential function to derive rising and decaying time constants, and show the best-fit model in the figure. 
%to extract the decay time constant. An example of a fit result is shown in the left panel of Fig. \ref{fig:tau_Tb}. 
The right panel of Fig. \ref{fig:tau_Tb} shows the best-fit decaying time constant as a function of the bias point 
%at $T_{\rm b}$ = 35 mK 
for all devices tested. A minimum is observed at a bias of approximately 10\%~$R_{\mathrm{n}}$. 

\begin{figure}[H]
\centering
\includegraphics[width=0.7\linewidth]{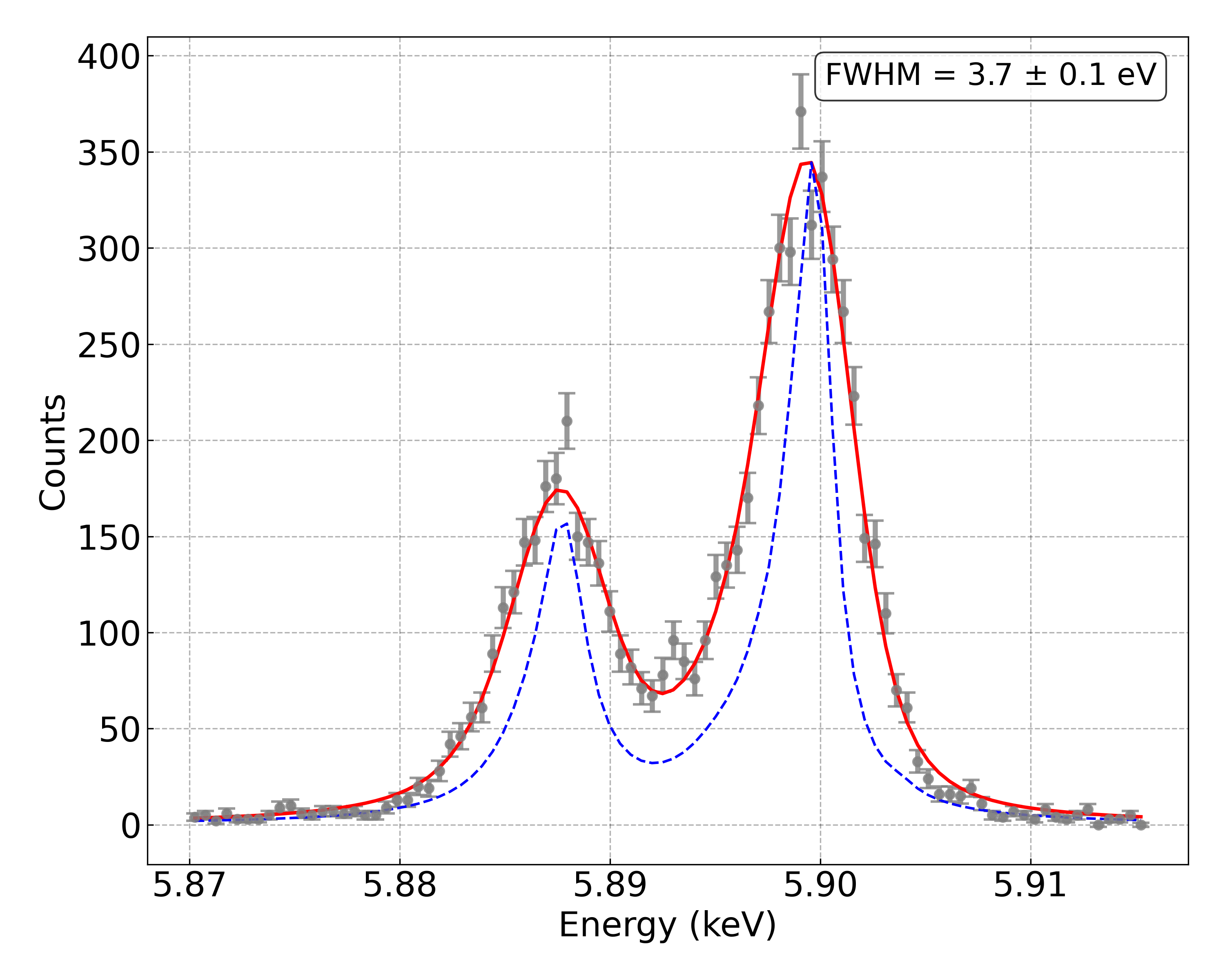}% Here is how to import EPS art
\caption{\label{fig:fit_kalpha} The measured X-ray spectrum of Device S2-5 around the Mn $\rm K_{\alpha}$ line. The doublet is clearly resolved. The best-fit model is shown in solid (red) line, with the natural line profiles shown in dashed (blue) line. 
}
\end{figure}

Again, pulse amplitudes are obtained with phase-compensated optimal filtering and are calibrated with the known energies of the Mn $\rm K_{\alpha}$ and $\rm K_{\beta}$ lines. Fig. \ref{fig:fit_kalpha} shows the calibrated X-ray spectrum around the $\rm K_{\alpha}$ line. The expected doublet is clearly resolved into $\rm K_{\alpha1}$ and $\rm K_{\alpha2}$.  
The energy resolution of a device is derived from simultaneously fitting all the lines (including $\rm K_{\beta}$) with a series of Lorentzian functions (representing the natural profiles of the lines \cite{Mnkab}) that are convolved with a Gaussian function (representing instrumental broadening). Among the devices tested, 
Device S2-5 shows the best resolution, $3.7\pm 0.1$ eV (FWHM) at 5.9 keV, as shown in Fig. \ref{fig:fit_kalpha}, when it is biased at $\sim$1.5\% $R_{\rm n}$ and operated at a base temperature of 20 mK.

\begin{figure}[H]
    \centering
    \includegraphics[width=0.7\linewidth]{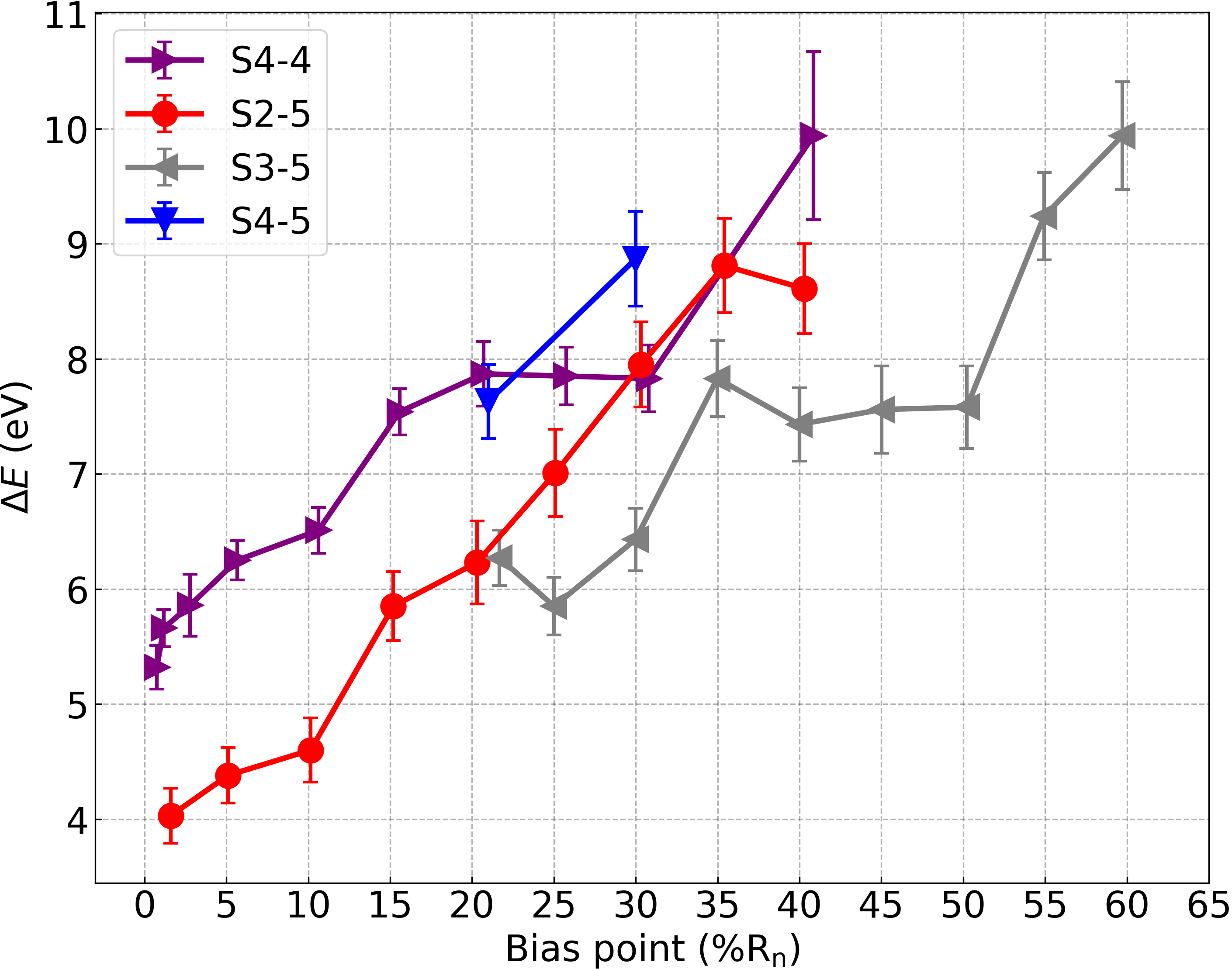}
    \caption{The measured energy resolution of the microcalorimeters at different biases. 
    }
    \label{fig:resolution_bias}
\end{figure}

We also measured the energy resolution of each device at the base temperature of $35~\mathrm{mK}$, and the results are shown in Fig. \ref{fig:resolution_bias}. 
All of our devices exhibit degradation in energy resolution towards high bias, with no apparent turning point at low bias, suggesting the absence of the so-called ``excess noise'' (e.g., \cite{fraser2004nature}, \cite{brandt2010analytical}). We note that no measurement with the pulsed laser system is attempted on these devices because they can not resolve 3 eV photons.

\subsection{Noise Spectrum Measurement} \label{sec:xnoise}
\begin{figure}[H]
    \centering
    \includegraphics[width=0.7\linewidth]{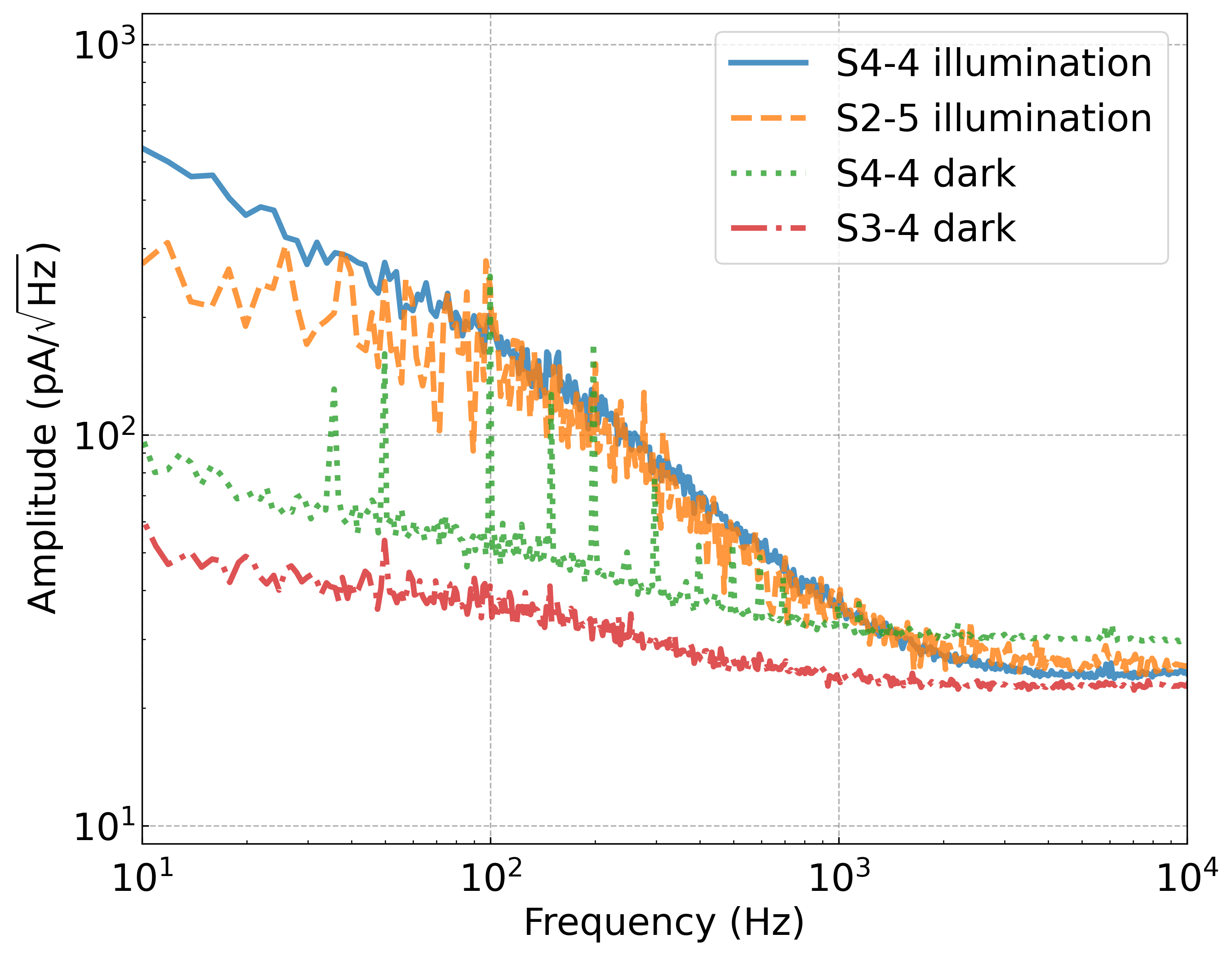}
    \caption{
    Current noise spectra of the microcalorimeters. The results are shown both for the illuminated devices and those in dark for direct comparison.
    }
    \label{fig:Xtalk}
\end{figure}

Fig. \ref{fig:Xtalk} shows the current noise spectra of Devices S4-4 and S2-5 that are obtained under illumination by the $^{55}$Fe source, when biased at  5\% $R_{\rm n}$ and operated at $35~\mathrm{mK}$. Each noise spectrum is computed using data in the segments of the time series that contain no apparent pulses. Also shown are the noise spectra of Devices S4-4 and S3-4 that are kept in the dark, when biased at the same level and operated at the same temperature for direct comparison. Note that the dark noise is higher for Device S4-4; this is due to a known issue with RF shielding, which is corrected for Device S3-4. 

It is evident that the noise is significantly higher for the illuminated devices than for the shielded devices, at frequencies up to roughly $1~\mathrm{kHz}$, 
which corresponds to the thermal response bandwidth of the devices.
This difference is very likely due to thermal crosstalk.
When a neighboring pixel absorbs an X-ray photon, the heat generated propagates through the substrate to the pixel under test, thus introducing additional thermal fluctuation noise.
The fundamental limit on the energy resolution of a microcalorimeter is set by the noise equivalent power (NEP) \cite{mccammon2005thermal}:
%is given by the standard expression from optimal filtering theory 
\begin{equation}
\Delta E_{\mathrm{FWHM}} = 2\sqrt{2\ln 2} \left( \int_{0}^{\infty} \frac{4}{|\mathrm{NEP}(f)|^{2}} {\rm d}f \right)^{-1/2}.
\label{eq:deltaE_nep}
\end{equation}
The thermal crosstalk raises the NEP level of an illuminated device, degrading its energy resolution.

\subsection{Complex Impedance Measurements} \label{sec:impedance}
\begin{figure}[H]
    \centering
    \includegraphics[width=0.5\linewidth]{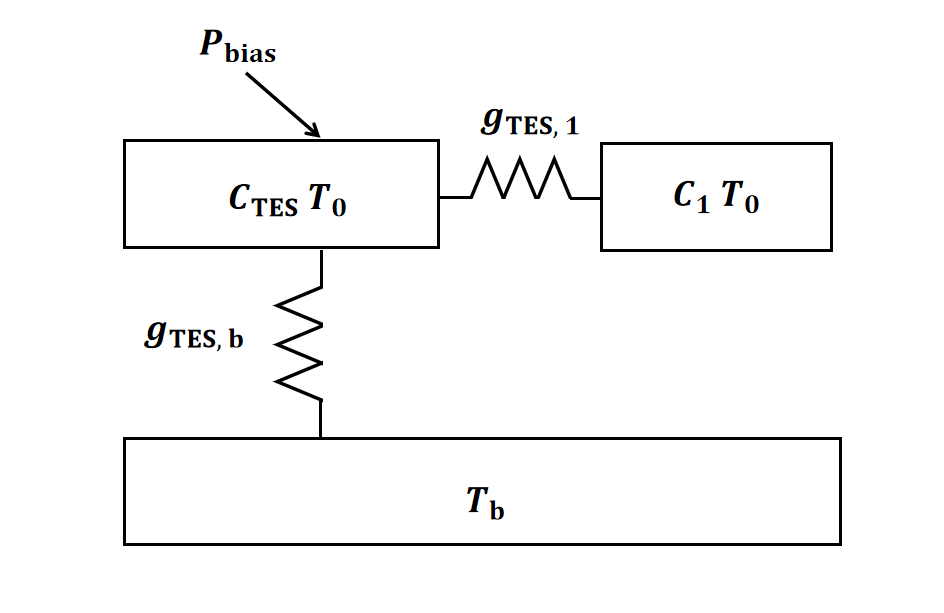}
    \caption{
   Two-block hanging device model.
    }
    \label{fig:IHmodel}
\end{figure}

Complex impedance analysis is a useful tool for deriving the electro-thermal properties of a microcalorimeter \cite{maasilta2012complex}, which are otherwise difficult to obtain. The measurements are made by adding a small sinusoidal perturbation to the bias voltage of a device over a selected range of frequencies (e.g., from 10 Hz to 10 kHz) and recording its current response.
%The complex impedance data were obtained by applying a series of small-amplitude sinusoidal signals at various frequencies (from 10 Hz to 10 kHz) around a given bias point.
We fit the complex impedance data of the microcalorimeters using a 2-body hanging model \cite{maasilta2012complex}, as illustrated in Fig. \ref{fig:IHmodel}. The model contains two heat capacities, with
$C_1$ representing the absorber, and $C_\mathrm{TES}$ the TES.

We adopt a differential evolution approach  \cite{helenius2020simultaneous}, to perform a simultaneous fitting of complex impedance and noise spectra. 
% The data acquired at bias points of 5\%, 10\%, and 15\% $R_{\rm n}$ are jointly fitted with the model, under the assumption that the parameters $C_1$, $C_2$, $g_\mathrm{TES,1}$, $g_\mathrm{TES, 2}$ and $T_2$ are independent of the device bias, while $\alpha_i$, $\beta_i$ and $C_{\mathrm{TES},i}$ are (with the subscript $i$ denoting the $i$-th bias point). 
%The thermal conductance $g_\mathrm{2, b}$ was derived from $g_\mathrm{eff}$ and $g_\mathrm{TES,2}$,
% The conductance $g_{\mathrm{eff}}$ obtained from an $I$-$V$ curve corresponds to the effective conductance of $g_{\mathrm{TES},2}$ and $g_{2,\mathrm{b}}$ that are connected in series, allowing $g_{2,\mathrm{b}}$ to be derived.
The values of $T_\mathrm{0}$, $P_{\rm bias}$ and $g_{\mathrm{TES, b}}$ are taken directly from the $I$-$V$ measurements. 
Our fit assumes that $g_{\mathrm{TES, 1}}$ and $C_1$ are common to all bias points, while $\alpha$, $\beta$, and $C_{\mathrm{TES}}$ differ between bias points. 

% In total, therefore, the model contains 14 free parameters. By constraining the bias-independent parameters across multiple bias points, the simultaneous fitting strategy narrows their allowed range of variation and yields more reliable and accurate results.
% In the mid-frequency range, the complex impedance fit is not fully satisfactory, suggesting that a higher-order model may be needed for a perfect description. To avoid overfitting, however, we do not pursue this further here. 
% Once the common parameters are determined, $\alpha$ and $\beta$ for the remaining bias points are obtained by fitting the complex impedance alone, as the thermal fluctuation noise (TFN) at these bias points is no longer significant compared with the readout noise.

\begin{figure}[H]
    \centering
    \includegraphics[width=0.65\linewidth]{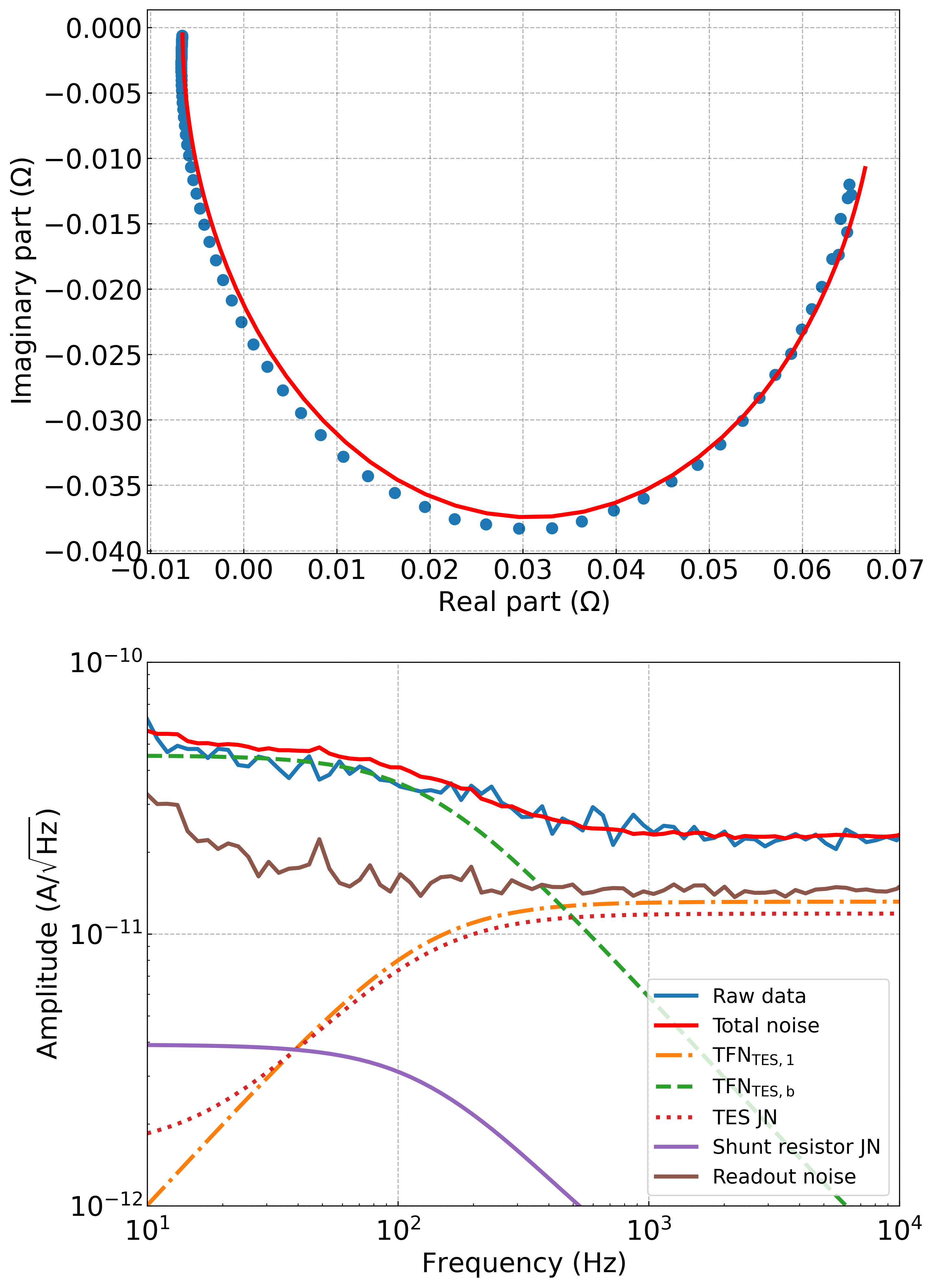}
    \caption{The complex impedance (top) and current noise spectrum (bottom) of Device S3-4 measured at the bias of 5\% $R_{\rm n}$.
    The best-fit model is shown in solid (red) line. 
    The decomposition of the noise spectrum is shown in the bottom panel with each contribution indicated. 
    % Note that the heat exchange between the substrate and $C_2$ (TFN$_{\rm b,2}$) stays below $10^{-12}~\mathrm{A}/\sqrt{\mathrm{Hz}}$ and is not displayed.
    }
    \label{fig:Complex}
\end{figure}

\begin{figure}[ht]
    \centering
    \includegraphics[width=0.6\linewidth]{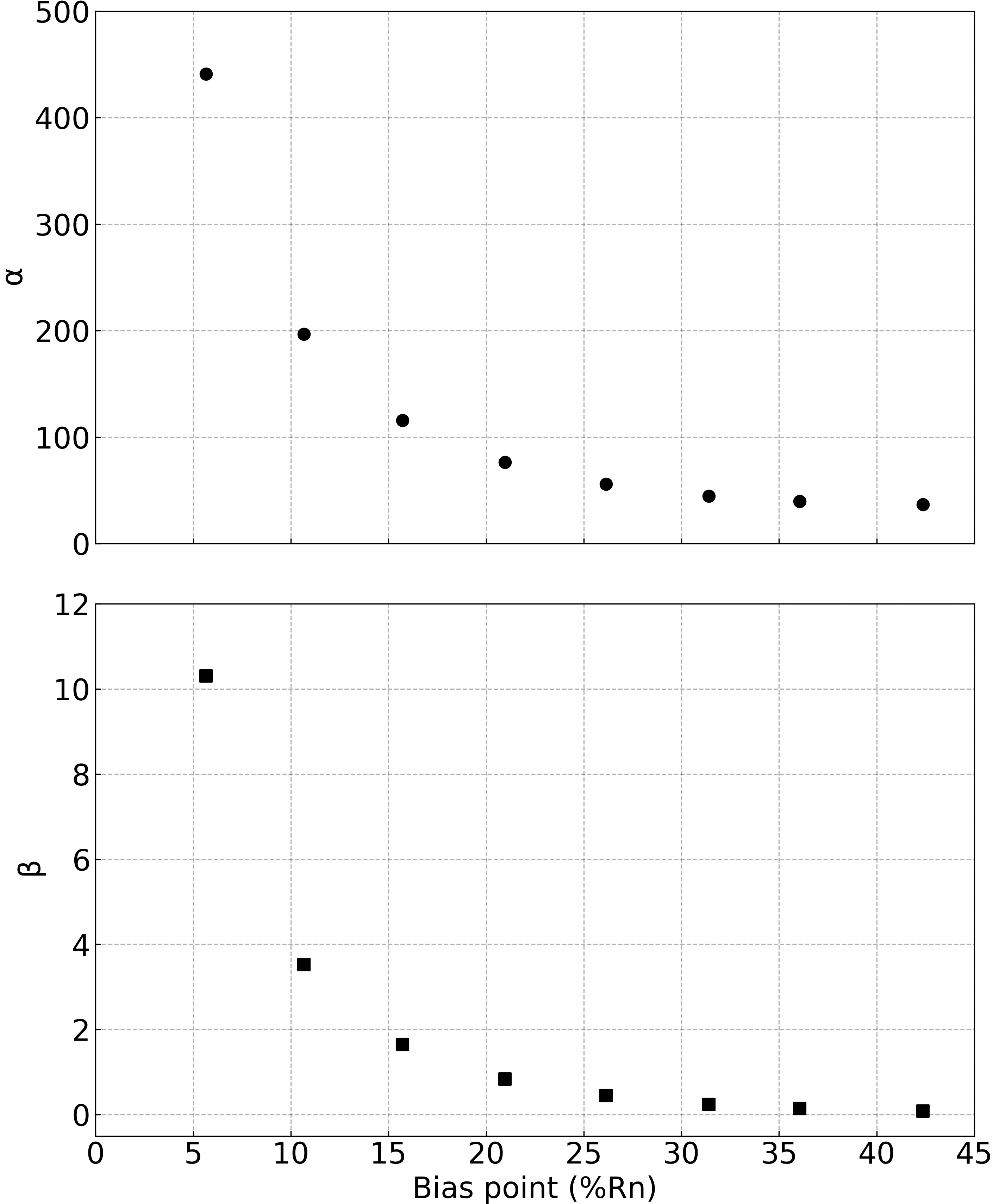}
    \caption{
    The temperature sensitivity $\alpha$ and current sensitivity $\beta$. Both were fitted from the complex impedance data at different bias points.
    }
    \label{fig:alpha_beta_bias}
\end{figure}

Fig. \ref{fig:Complex} shows the measured complex impedance and noise spectrum of Device S3-4, as an example, when it is biased at 5\% $R_{\rm n}$. The results of simultaneously fitting the two sets of data with the model are also shown in the figure. In the mid-frequency range, the complex impedance fit is not fully satisfactory, suggesting that a higher-order model may be needed for a perfect description. To avoid overfitting, however, we do not pursue this further here. 
The fitted value of $C_1$ is $0.77\,\mathrm{pJ/K}$, and that of $g_{\mathrm{TES}, 1}$ is $110\,\mathrm{nW/K}$.
During the fitting, we found that $C_{\mathrm{TES}}$ is almost unconstrained by the model, and its values are therefore not listed here.
The fitting results for $\alpha$ and $\beta$ are shown in Fig. \ref{fig:alpha_beta_bias}.

The results suggest that the thermal fluctuation noise associated with heat exchange between the TES and heat sink (TFN$_{\rm TES, b}$) dominates at low frequencies, while the thermal fluctuation noise associated with heat exchange between the TES and $C_1$ (TFN$_{\rm TES, 1}$) and the Johnson noise (JN) of the TES together account for the high-frequency noise (after considering the readout noise).
Again, there is no evidence for contribution from the    ``excess noise'' here.

% \begin{table}[htbp]
%     \centering
%     \caption{Value of common parameters in complex impedance fitting.}
%     \label{tab:complex_params}
%     \begin{tabular}{@{}c c c c c@{}}  % @{} 消除两侧多余空白
%         \toprule
%         $C_1~\mathrm{(fJ/K)}$ & $C_2~\mathrm{(fJ/K)}$ & $T_2~\mathrm{(mK)}$ &  $g_\mathrm{TES,1}~\mathrm{(nW/K)}$ &  $g_\mathrm{TES, 2}~\mathrm{(pW/K)}$ \\
%         \midrule
%         $619 \pm 5$ & $134 \pm 5$ & $35.8 \pm 0.3$ & $136 \pm 9$ & $81 \pm 4$ \\

%         \bottomrule
%     \end{tabular}
% \end{table}

\section{Discussion}

\subsection{Energy Resolution as a Function of Bias Point}
\begin{figure}[H]
\centering
\includegraphics[width=0.7\linewidth]{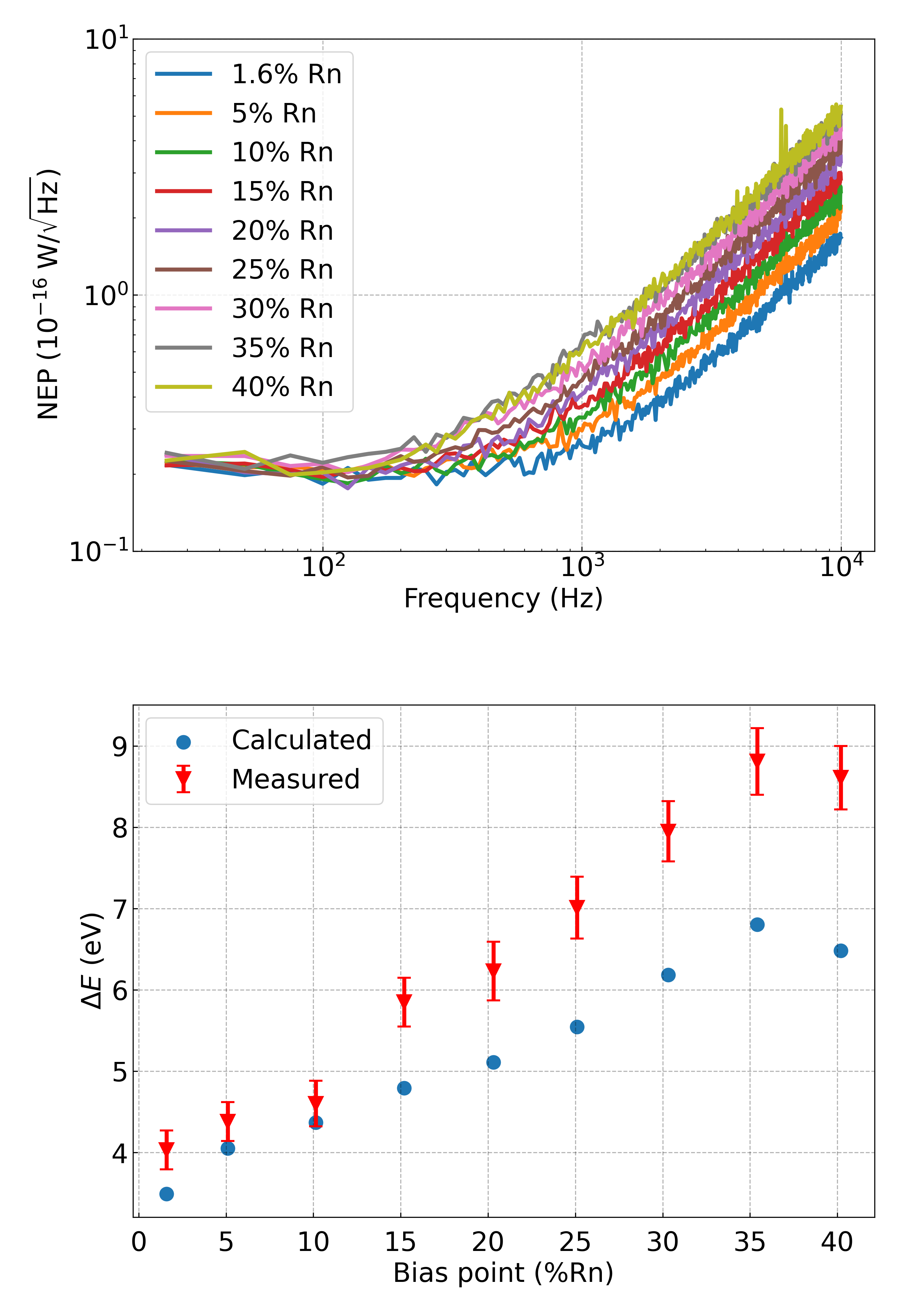}% Here is how to import EPS art
\caption{\label{fig:NEP_vs_dE} The NEP of device S2‑5 at 35 mK and a comparison of the calculated and measured resolutions. Top: NEP spectra (25 Hz–10 kHz) at nine bias points from 1.6\% Rn to 40\% Rn. Bottom: comparison between the NEP‑integrated calculated resolution (circles) and the measured resolution (triangles).
}
\end{figure}

As shown in Fig. \ref{fig:resolution_bias}, the energy resolution of all devices exhibits a degradation trend with increasing bias point. Taking device S2‑5 as an example, we found that the energy resolution roughly follows
%is positively correlated with the decreasing trend of 
1/$\sqrt{\alpha}$, as in Eq.~\ref{Eresolution}.
%(assuming that the sensitivity performance of all devices in this batch is close to that shown in Fig.\ref{fig:alpha_beta_bias}), suggesting that $\alpha$ is the dominant parameter governing the energy resolution of the devices.
To investigate the trend further, we calculate the NEP of the device at different bias points, and compare the expected energy resolution with the measured values. In the frequency domain, the NEP follows directly as $\text{NEP}(f) = \sqrt{S_I(f)}/|s_I(f)|$, where $S_I$ is the current noise power spectral density of the device, and $s_I$ represents its small-signal current responsivity, which
%$s_I(f) = dI/dP$ was
is obtained with the Fourier transform of the average X-ray pulse,
%\cite{taralli2021performance}, 
%. The responsivity was then computed as
$s_I(f) = \mathcal{F}[\langle I(t)\rangle]/E$, where $\langle I\rangle$ is the average pulse profile in time domain, and $E$ is the total energy in the pulse ($E = 9.44 \times 10^{-16}\ \text{J}$ in our case). $S_I(f)$ is computed from 
%40 ms segments of 
pre-trigger data,
%acquired under X-ray illumination, 
using the same record length and sampling rate as the pulse data. The results are shown in Fig. \ref{fig:NEP_vs_dE} for different bias points. 

From $\text{NEP}(f)$, the expected energy resolution is calculated with Eq.~\ref{eq:deltaE_nep} (with the integration limits set at 25 Hz and 10000 Hz). The results are also shown in Fig. \ref{fig:NEP_vs_dE}.
At frequencies below 200\,Hz, the NEP levels are nearly identical across all bias points, where the noise is dominated by TFN between TES and thermal bath. At higher frequencies, however, the NEP exhibits a clear increasing trend with rising bias point, which is primarily attributed to readout noise, followed by Johnson noise \cite{maasilta2012complex} (see the noise decomposition results in Fig. \ref{fig:Complex} for details). Therefore, the noise at high frequencies causes the degradation of the energy resolution with increasing bias. The lower panel of Fig. \ref{fig:NEP_vs_dE} shows a comparison between the calculated and measured energy resolutions as a function of bias point. Both show the same trend, but the calculated values are always below the measured ones, which is perhaps related to the fact that we have neglected contribution from noise at very low frequencies (below 25 Hz) in the calculation. 
%very closely for the first three bias points.
%Further reducing the readout noise could be an effective way to alleviate this degradation trend.

%However, starting from the 15\%\,Rn bias point, the resolution exhibits an overall degradation trend, which is puzzling. We suspected that this might be caused by a degradation of the electrothermal feedback (ETF) at this bias point, so we calculated the effective loop gain $\mathcal{L}_{\mathrm{eff}}$. The value of $\mathcal{L}_{\mathrm{eff}}$ is derived from the time constant $\tau_{\mathrm{eff}}$ (See Fig. \ref{fig:tau_Tb}) of the pulse falling edge via $\mathcal{L}_{\mathrm{eff}} = \tau_0/\tau_{\mathrm{eff}} - 1$, where $\tau_0$ is determined by the ratio of the heat capacity to the thermal conductance of the device. The results of $\mathcal{L}_{\mathrm{eff}}$ are presented in Fig. \ref{fig:L_eff}. It can be seen that $\mathcal{L}_{\mathrm{eff}}$ shows a slight downward trend at the 15\% bias point, but its absolute value does not change significantly, which is insufficient to indicate an obvious weakening of ETF. Therefore, the degradation trend in resolution caused by this mechanism is ruled out. The specific cause of this abrupt change may be related to variations in the external testing environment, such as magnetic field drift.
%\begin{figure}[H]
%\centering
%\includegraphics[width=0.7\linewidth]{14_LG_eff_vs_R.png}% Here is how to import EPS art
%\caption{\label{fig:L_eff} The $\mathcal{L}_{\mathrm{eff}}$ of device S2‑5 v.s. bias point.
%}
%\end{figure}

\subsection{Thermal Crosstalk in Illumination Test}
\begin{figure}[ht]
\centering
\includegraphics[width=0.6\linewidth]{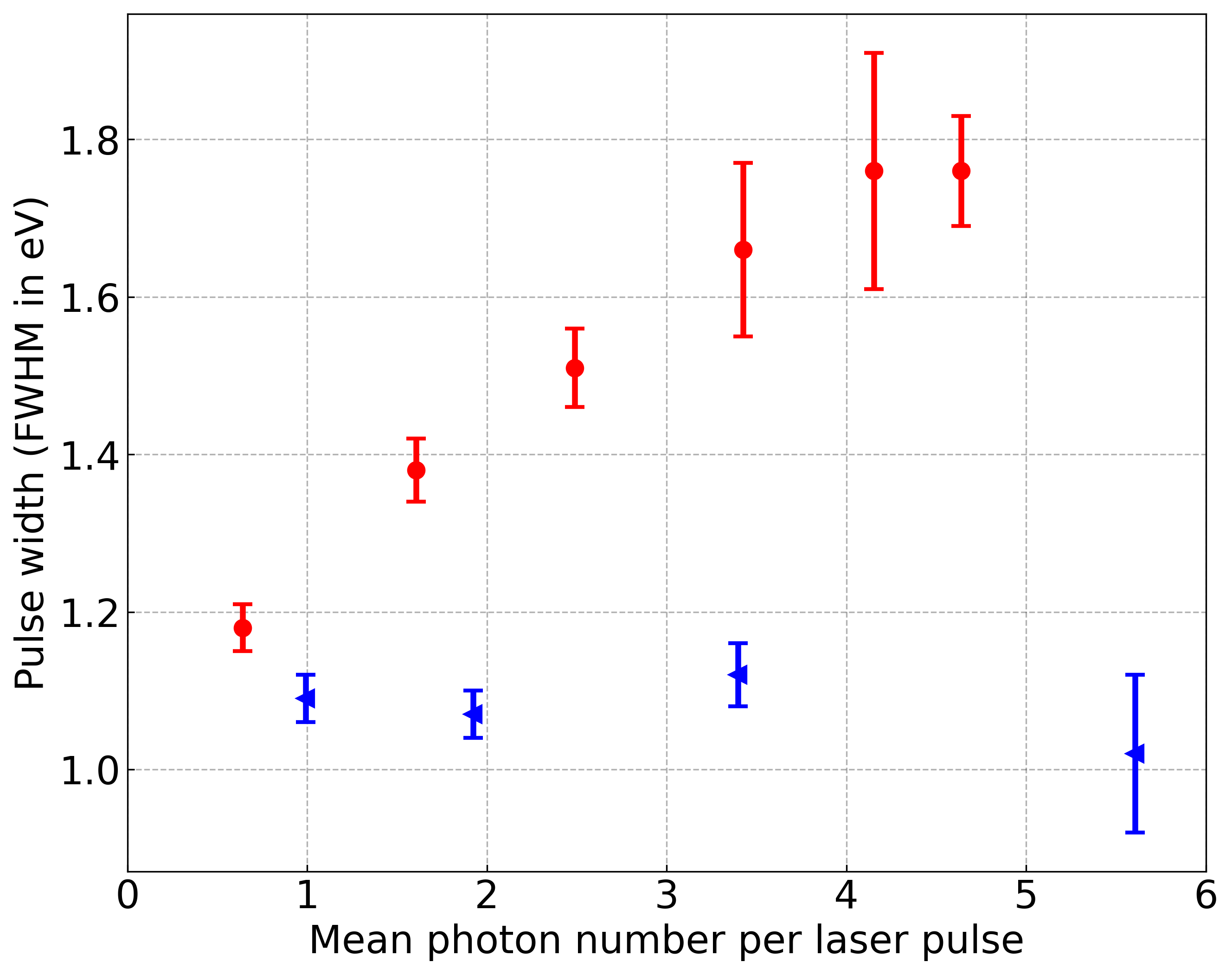}% Here is how to import EPS art
\caption{\label{fig:area} Effects for thermal crosstalk in optical illumination experiments. The measured widths of single-photon peaks are shown, as a function of laser power, for two setups: one with a pinhole of 3.14 times the TES area (in red dots) and the other with a collimating pinhole of 0.52 times the TES area (in blue triangles).
See the main text.
}
\end{figure}

In the pulsed-laser illumination experiment, a mask with a small pinhole is used to align the laser spot onto the TES, to prevent the photons from hitting the SiN membrane surrounding the TES, which causes significant thermal crosstalk, as shown in Fig. \ref{fig:area}. In the figure, the widths of single‑photon peaks are compared between two sets of measurements, the red dots correspond to a device (denoted TES1) for which the ratio of the pinhole area to the TES area is $3.14$, whereas the blue triangles represent the device already shown in Fig. \ref{fig:Eresolution} (denoted TES2), for which the area ratio is $0.5$ (and the pinhole diameter equals the short side of the TES). As the laser power is adjusted (quantified by the mean photon number per pulse), the energy resolution of TES1 degrades significantly toward high laser power, while that of TES2 remains roughly constant. We attribute the degradation in energy resolution to the thermal crosstalk caused by photons depositing energy in the SiN membrane. Similar effects have been seen before, e.g., in \cite{akamatsu2009}, who interpreted them as excess phonon-like noise.
On the other hand, Fig.~\ref{fig:Eresolution} shows that, at a fixed laser power, the energy resolution degrades significantly as the photon number increases, even with a small pinhole (cf.~\cite{jaeckel2021calibration}). 
We find that the trend can be well described with the downconversion phonon noise model\cite{kozorezov2006resolution}, which implies $\Delta E = 2\sqrt{2\ln 2}\,\sqrt{\Delta E_0^2 + J \cdot E}$, where $J$ is the downconversion phonon noise factor and $\Delta E_0$ denotes the baseline energy resolution. By fitting our data with this model, we obtain $J \approx 0.015$ eV. This value is one order of magnitude larger than $J \approx 0.001$ eV reported for W/Si optical TESs \cite{kozorezov2006resolution}, perhaps suggesting that athermal phonon loss is substantially stronger in our Mo/Cu films on a solid SiN/Si substrate, likely due to a stronger electron-phonon coupling. The trend of the degradation in energy resolution is also much steeper than in other recent works \cite{jaeckel2021calibration, roy2026characterization}. Fabricating the TES on a suspended SiN membrane is expected to reduce this effect via reflecting escaping phonons back into the TES and therefore improve the energy resolution \cite{lita2008, kozorezov2013athermal}.

In X-ray illumination of microcalorimeters, thermal crosstalk may also occur. In this case, the SiN membrane is almost entirely covered by the absorber, so the membrane-related thermal crosstalk is likely negligible. 
Unlike the optical illumination experiment, however, X-ray photons irradiate all pixels, so thermal crosstalk due to neighboring pixels could be significant. Such effects manifest themselves in the elevated noise levels of the pixels that are illuminated by X-rays (as shown in Fig. \ref{fig:Xtalk}). To quantify the effects on the energy resolution, 
we convert the current noise $S_I(\omega)$ to $\mathrm{NEP}$, the energy resolution is then obtained via NEP integration. In the absence of optical pulses, the current response $s_I(\omega)$ can be obtained from the complex impedance  \cite{maasilta2012complex}, 
% \cite{irwin2005transition}:
% \begin{equation}
% \mathrm{NEP^2}= S_I(\omega)/s_I(\omega),
% \label{eq:deltaE_nep_sq}
% \end{equation}
%where $s_I(\omega) = -\frac{1}{Z_{\mathrm{circ}} I_0} \frac{Z_{\mathrm{TES}} - R_0(1 + \beta)}{R_0(2 + \beta)}$,
where $s_I(\omega) = - \bigl[ 1 / (Z_{\mathrm{circ}} I_0) \bigr] \cdot \bigl[ (Z_{\mathrm{TES}} - R_0(1 + \beta)) / (R_0(2 + \beta)) \bigr]$, 
$Z_{\mathrm{circ}} = Z_{\mathrm{TES}} + R_L + i\omega L$. 
Applying it to the noise spectrum of Device S3-4 (shielded from X-ray illumination; see Tab. \ref{tab:ucal}), from Eq. \ref{eq:deltaE_nep}, we obtain a calculated energy resolution of 1.6 eV, at a bias point of 5\% $R_{\rm n}$, in the absence of illumination. For Device S2-5 (under X-ray illumination), we get a calculated energy resolution of 4.0 eV at 5\% $R_{\rm n}$ , in the presence of illumination. Assuming that the devices are intrinsically similar,  
we estimate that the degradation in energy resolution observed under illumination reaches approximately 3.7 eV, which is most likely caused by thermal crosstalk.

A more accurate quantification of the effects of thermal crosstalk can be achieved by simultaneously reading out two adjacent pixels.
Such measurements will be pursued in future work. To alleviate the effects, we plan to improve the thermal conductance between the TES and the heat sink, following \cite{Xtalk_heatsink1}.

\section{Conclusion}
We have fabricated TES and microcalorimeter arrays, based on Mo/Cu bilayer films, and measured their electrothermal properties. The results confirm that the designed TES contributes little to the energy resolution of the TES microcalorimeter. At the energies of Mn K$_\alpha$ lines ($\sim$5.9 keV), the best energy resolution obtained is 3.7$\pm$0.1 eV among the tested microcalorimeters. Comparing the noise spectra of the illuminated and shielded devices, we have identified pixel-to-pixel thermal crosstalk as a significant factor that degrades the energy resolution in our experimental setup. This is an area for improvement in the future.

Development is also underway to replace small-area pure Au absorbers with large-area Au/Bi ones, which are baselined for HUBS. It would, in principle, maintain a similar energy resolution since the heat capacity is designed to be the same. However, the reduction of Au thickness will probably cause a degradation of energy resolution due to the position dependence effect on the pulse \cite{wang2025modeling}, which needs further investigation. Finally, efforts will be made to improve the uniformity of pixels within an array, between arrays in the same batch, and between different batches.

\section*{Acknowledgement}
We wish to thank Dr. Jingjing Li and her team at the National Institute of Metrology for providing the input SQUIDs used in our experiment, and Dr. Hiroki Akamatsu for helpful comments on the manuscript. Sifan Wang wishes to acknowledge support from the China Postdoctoral Science Foundation through Grant GZB20240377. This work was supported in part by the National Natural Science Foundation of China through Grant 12220101004, by the Ministry of Science and Technology of China through Grant 2022YFC2205100, and by the China National Space Administration through a technology development grant.




\bibliographystyle{elsarticle-num} 
\bibliography{HUBS1-TESmuCal-ref}


% %% For numbered reference style

%   \textit{\LaTeX: a document preparation system},
%   Addison Wesley, Massachusetts,

% \end{thebibliography}

% \appendix
% \section{Appendix}
% % Appendix text.

\appendix
\section{Derivation of the Phase-Compensated Optimal Filter}
\label{app1}
The standard OF rests on two assumptions: the detector response is linear (all pulses share a
common shape $s(t)$ and differ only in amplitude $E$) and the noise is stationary and Gaussian.
In the frequency domain, a recorded pulse is modelled as
\begin{equation}
D(f_k) = E\,S(f_k) + N(f_k),    
\end{equation}
where $S(f_k)$ is the Discrete Fourier Transform (DFT) of the normalised pulse template and $N(f_k)$ is the noise.  The
optimal amplitude is found by minimising the noise-weighted sum of squared residuals,
\begin{equation}
\chi^2(E) = \sum_k \frac{1}{J(f_k)}\;
\bigl| D(f_k) - E\,S(f_k) \bigr|^2 \,\Delta f,
\qquad J(f_k) \equiv \langle |N(f_k)|^2 \rangle,
\end{equation}
yielding the classical frequency-domain OF estimator,
\begin{equation}
\hat{E} = \frac{\displaystyle
\sum_k \frac{1}{J(f_k)}\;
\operatorname{Re}\!\bigl[ D(f_k)\,S^{\ast}(f_k) \bigr]\,\Delta f}
{\displaystyle
\sum_k \frac{1}{J(f_k)}\;|S(f_k)|^2\,\Delta f}.\label{eq:OF_old}
\end{equation}
Equivalently, in the time domain, $\hat{E} = \kappa\sum_i d(t_i)\,\mathrm{OF}(t_i)$, where
$\mathrm{OF}(t) \equiv \mathcal{F}^{-1}[S^{\ast}(f)/J(f)]$ is the optimal filter kernel.
When the photon arrival time is random with respect to the sampling clock, the pulse acquires a
sub-sample time offset $\tau$:

\begin{equation}
d(t_i) = E\,s(t_i - \tau_0) + n(t_i).\label{eq:time_domain} 
\end{equation}
In the frequency domain, by the shift theorem,
\begin{equation}
D(f_k) = E\,S(f_k)\,e^{-j\omega_k\tau_0} + N(f_k), \qquad \omega_k \equiv 2\pi f_k.\label{eq:phase_shift}    
\end{equation}
At the experimental sampling rate of $100\;\text{kHz}$ (after anti-aliasing), this random arrival phase degrades the energy resolution if left uncorrected.

Following \cite{szymkowiak1993signal}, the time offset can be recovered by convolving the pulse
with the optimal filter kernel and identifying the lag that maximises the output. Defining

\begin{equation}
\mathcal{C}(\tau) \equiv \sum_i d(t_i)\,\mathrm{OF}(\tau - t_i)    
\end{equation}
and inserting $d(t_i) = E\,s(t_i - \tau_0) + n(t_i)$,

\begin{equation}
\mathcal{C}(\tau) = E \underbrace{\sum_i s(t_i - \tau_0)\,\mathrm{OF}(\tau - t_i)}_{\text{signal}}
                  \;+\; \underbrace{\sum_i n(t_i)\,\mathrm{OF}(\tau - t_i)}_{\text{noise}}.   
\end{equation}
The noise term has zero expectation, $\langle\sum_i n\cdot\mathrm{OF}\rangle = 0$, because
$\langle n\rangle = 0$.  The signal term is the cross-correlation of the pulse shape with a
matched filter constructed from $S^{\ast}(f)/J(f)$; it attains a unique maximum when the two
are perfectly aligned, i.e.\ $\tau = \tau_0$.  Hence

\begin{equation}
\hat{\tau} = \underset{\tau}{\arg\max}\;|\mathcal{C}(\tau)|.
\end{equation}

To achieve sub-sample precision, the cross-spectrum $D(f_k)S^{\ast}(f_k)/J(f_k)$ is zero-padded
by an integer factor $u$ (typically $u=10$) before the inverse DFT.  This corresponds to sinc
interpolation of the time-domain cross-correlation sequence and evaluates
$\mathcal{C}(\tau)$ on a grid with effective spacing $\Delta t / u$.  The peak location is further
refined by a three-point parabolic fit.  Let $m_0 = \arg\max_m|\mathcal{C}_m|$ and
$a = |\mathcal{C}_{m_0-1}|$, $b = |\mathcal{C}_{m_0}|$, $c = |\mathcal{C}_{m_0+1}|$; the sub-bin
correction is

\begin{equation}
\delta = \frac{1}{2}\,\frac{a - c}{a - 2b + c}.
\end{equation}

The final estimate in units of the sampling interval is
$\hat{\tau} = (m_0 - uN/2 + \delta)/u$, restricted to $[-0.5,+0.5]$. With $\hat{\tau}$ known, the amplitude follows from Eq.~\eqref{eq:OF_old} with the template shifted to
cancel the phase ramp in Eq.~\eqref{eq:phase_shift}:
\begin{equation}
\hat{E} = \frac{\displaystyle
\sum_k \frac{1}{J(f_k)}\;
\operatorname{Re}\!\bigl[ D(f_k)\,S^{\ast}(f_k)\,e^{+j\omega_k\hat{\tau}} \bigr]\,\Delta f}
{\displaystyle
\sum_k \frac{1}{J(f_k)}\;|S(f_k)|^2\,\Delta f}.
\label{eq:phase_corrected}  
\end{equation}
The factor $e^{+j\omega_k\hat{\tau}}$ cancels the $e^{-j\omega_k\tau_0}$ ramp on $D(f_k)$, so that the
product $D(f_k)S^{\ast}(f_k)e^{+j\omega_k\hat{\tau}}$ becomes, modulo residual noise, purely real.
When $\hat{\tau} \to 0$, Eq.~\eqref{eq:phase_corrected} reduces to the standard OF estimator Eq.~\eqref{eq:OF_old}.

\end{document}